\documentclass[letterpaper,twocolumn,10pt]{article}
\usepackage{usenix,epsfig,endnotes}

\usepackage[utf8]{inputenc}
\usepackage{graphics}
\usepackage{graphicx}
\usepackage{cite}
\usepackage{xcolor}
\usepackage[]{algorithm2e}
\usepackage{float}
\usepackage[bookmarks=true,%
            bookmarksnumbered=true,%
            colorlinks=true,%
            linkcolor=linkcol,%
            citecolor=citecol,%
            urlcolor=urlcol,%
            hypertexnames=true,%
            pdfpagelabels]%
      {hyperref}
   \hypersetup{ pdfauthor = {Gianluca Stringhini},
                pdftitle = {},
                pdfkeywords = {},
                pdfcreator = {},
                pdfproducer = {}
              }

   \definecolor{linkcol}{rgb}{0,0,0.5}
   \definecolor{citecol}{rgb}{0,0.5,0.3}
   \definecolor{urlcol}{rgb}{0.3,0,0}
\usepackage{url}
\usepackage{hyperref}
\usepackage{todonotes}
\usepackage{amsfonts,pifont}
\usepackage{amstext}
\usepackage[font=footnotesize]{subfig}
\usepackage{fixltx2e}
\usepackage{flushend}

\begin{document}

\title{Honey Sheets: What Happens To Leaked\\ Google Spreadsheets?}

\author{
	{\rm Martin Lazarov, Jeremiah Onaolapo, and Gianluca Stringhini} \\
	University College London \\
	martin.lazarov.12@ucl.ac.uk,\{j.onaolapo,g.stringhini\}@cs.ucl.ac.uk
}

\maketitle

\subsection*{Abstract}
Cloud-based documents are inherently valuable, due to the volume and nature 
of sensitive personal and business content stored in them. 
Despite the importance of such documents 
to Internet users, there are still large gaps in the 
understanding of what cybercriminals do when they illicitly get access to them
by for example compromising the account credentials they are associated with. 
In this paper, we present a system able to monitor user activity on Google
spreadsheets.  
We populated 5 Google spreadsheets with 
fake bank account details and fake funds transfer links. Each spreadsheet 
was configured to report details of accesses and clicks on links back 
to us. To study how people interact with these spreadsheets in case they are
leaked, we posted unique links pointing to the spreadsheets on a popular 
paste site. We then monitored activity in the accounts for 72 days, 
and observed 165 accesses in total. We were able to observe interesting
modifications to these spreadsheets performed by illicit accesses. For 
instance, we observed deletion of some fake bank account information, in addition to 
insults and warnings that some visitors entered in some of the spreadsheets.
Our preliminary results show that our system can be used to shed light on
cybercriminal behavior with regards to leaked online documents.  

\section{Introduction}

Many useful services are cloud-based nowadays, for example 
Dropbox~\cite{dropboxcom} and OneDrive~\cite{onedrive} which are both file storage services. Reasons for 
using cloud-based storage services include ease of access to files 
stored in the cloud and the ability to collaborate easily on shared 
files. According to a study by Eurostat, 21\% of EU citizens stored 
documents and multimedia files in the cloud as of 2014~\cite{eurostat:cloudstats}. 
Apart from individuals, many organizations also rely heavily on 
cloud-based services for office duties, such as Google Apps for Work~\cite{googleappsforwork}, 
a cloud-based suite that provides business tools like official 
email addresses, calendars, online storage space, and document 
processing services.  

Consequently, a considerable amount of valuable information is stored in online 
accounts, including sensitive personal information and business secrets. 
These in turn attract cybercriminals seeking to make profit from such 
information. 
To gain access to such online accounts, cybercriminals 
typically target owners of the accounts with phishing attacks, and 
sometimes, malware. Often, databases containing user credentials are 
also attacked, and credentials in them get stolen.

Stolen credentials are usually sold by cybercriminals in underground 
marketplaces, to other cybercriminals for malicious purposes. Depending 
on the nature of the compromised accounts, some are used to send 
unsolicited messages, and some are assessed by the 
cybercriminals with the hope of finding other valuable information 
that could be used to stage further attacks, either against the victim, 
or the victim's contacts. Such information of 
interest to cybercriminals include contacts lists, financial details 
and authentication credentials linked to other accounts.

Previous work on compromised cloud-based accounts is limited. They focus 
on malicious activity in compromised webmail accounts, with emphasis on 
spearphishing as a primary 
attack vector~\cite{bursztein2014handcrafted,stringhini2015ain}. It is 
hard for researchers to collect data on compromised accounts because 
there are no existing publicly available systems built for that purpose. 
To partially bridge this gap, we develop a system based on Google Apps
Script~\cite{googleappscript} to monitor compromised 
online spreadsheets.   
Our system is able to monitor activity performed by users on Google
spreadsheets, such as open events, modifications, and clicks on links.
Having visibility on these events is key for researchers wanting to understand
what is happening to cloud documents once they are accessed by attackers.

To demonstrate the usefulness of our system to study illicit accesses to
online documents, we set up honey spreadsheets in Google Docs and populate 
them with fake financial information. 
We also embed obfuscated links 
in the spreadsheets, some pointing to a domain under our control, others 
pointing to invalid pages on some selected banks' websites. We leak 
unique links pointing to the spreadsheets on a popular paste site, after 
configuring each spreadsheet to allow anyone with the links to edit the 
spreadsheets. 
We set up two scenarios on why these spreadsheets could be posted on paste sites: a
user innocently sharing them with her colleagues, inadvertently making it
accessible to anyone on the Internet, and a criminal posting a link to the
spreadsheet after having obtained illicit access to it. 
Compromised credentials and 
other sensitive information are routinely posted by cybercriminals 
on paste sites~\cite{scmagazine2015}, so this scenario is believable. 

After posting the links, we monitor accesses to the honey 
spreadsheets, and log interactions and activity in them. We aim to 
answer the following research questions:

\noindent \textit{Question 1:} What actions do attackers take on stolen 
cloud-based documents? Can we identify typologies of cybercriminals 
based on the actions they take in the documents?
 
\noindent \textit{Question 2:} What kind of content do cybercriminals 
find more attractive than others, within the same document? What kind 
of content do they interact with more than others?

Over a period of 72 days, we conducted two experiments and observed 165 
accesses, 28 modification events and 174 clicks on the embedded obfuscated 
links. The clicks originated from 35 different countries. The modifications 
we observed include deletion of fake payment information, entry of insults 
into a spreadsheet, and defacement of the same spreadsheet, thus rendering 
it unusable.

This paper makes the following contributions:
\begin{itemize}
 \item We develop a novel system to track activity in Google spreadsheets. 
 The system is able to observe file opening and modification of 
 content, in addition to tracking clicks on links contained in the 
 spreadsheets. The system also records IP addresses, request paths and 
 HTTP headers for clicks on the embedded links.

 \item We deployed 5 honey spreadsheets and leaked links pointing 
 to them on a popular paste site. We perform experiments for 
 72 days and measure accesses to the spreadsheets. During the 
 experiments, visitors modified and deleted fake payment information, 
 expanded some columns in the spreadsheets in order to get better 
 views of the data in those columns, and also replaced an obfuscated 
 link in a spreadsheet with a C++ code snippet. 

 \item We discuss how our approach could be used to set up different scenarios
   and give the research community a better understanding of the modus operandi
   of cybercriminals compromising stolen online documents.

\end{itemize}

\section{Background}
\label{sec:background}

\subsection{Cloud-based spreadsheets}

As stated earlier, cloud-based accounts are popular nowadays, and a large amount
of documents particularly are stored in cloud-based storage facilities. 
We focus our study on Google spreadsheets, particularly on actions taken 
by cybercriminals when they compromise the spreadsheets. We chose 
Google spreadsheets since the Google platform offers scripting 
functionality that allows us to boost the capability of the 
spreadsheets, to monitor activity in them. Our methodology can be applied 
to other cloud-based documents as well. In this section, we summarize 
the capabilities that spreadsheets provide, and focus on Google 
spreadsheets specifically.

When logged in to a Google account, users can create a new spreadsheet, 
or import an existing one. Each spreadsheet comprises cells arranged in 
rows and columns. Users can delete existing rows and columns of cells, 
move existing rows and columns around, and insert new ones as well.

For purposes of collaboration, users can share spreadsheets with other 
users usually by listing their email addresses, or by sharing a unique link pointing 
to the spreadsheet. Depending on permission settings specified by the 
document owner, collaborators can have view-only, comment-only, 
or edit access.      

Spreadsheet owners can extend the functionality of their spreadsheets by 
leveraging Google Apps Script~\cite{googleappscript}, a cloud-based 
Javascript engine for attaching scripts to documents in Google Docs platform. 
The scripts may be triggered by events such as opening of the spreadsheet or 
modification of its contents. Such scripts may also be triggered at some 
specified time intervals. Google Apps Script is commonly used for 
development of minimal web applications, in addition to 
extending the functionality of Gmail accounts and other Google 
products~\cite{googleappscript}. Google Apps Script can be configured to send 
out notifications when triggered, among other tasks.

\subsection{Threat model}

Cloud-based documents can get compromised in a number of ways. 
Cybercriminals mount phishing and spearphishing attacks on online 
accounts; this is the most common way by which cybercriminals manually hijack 
online accounts, according to Bursztein et al.~\cite{bursztein2014handcrafted}.
Information-stealing malware also capture login credentials safeguarding online 
accounts, for instance after drive-by downloads or clicking on spurious 
links and attachments in emails. Attackers also compromise accounts by 
hacking databases containing login credentials. 

Poor file access configuration, for example granting ``allow-all'' 
permissions, rather than specifying a whitelist of 
permitted parties, can also expose online documents to threats. 
Attackers have easy access if they get hold of pointers to a file, 
for instance by eavesdropping on communications about shared files,
and grabbing links pointing to such files.

Attackers can also mount Man-in-the-Cloud attacks on documents 
stored online, by stealing authentication tokens and impersonating 
legitimate parties that have access to the compromised documents~\cite{imperva2015}.   

Previous work focused on what happens in webmail 
accounts compromised via phishing 
attacks~\cite{bursztein2014handcrafted}. However, our work 
focuses on attacks on documents in the cloud. Similarly, Stolfo et al. 
proposed a method of mitigating insider threats to confidential data 
stored in the cloud~\cite{stolfo2012fog}. Their work, however, 
focuses on insider threats, while we consider a broader threat model.

\section{Methodology}
\label{sec:methodology}

\begin{figure}[th]
  \centering
  \includegraphics[width=0.45\textwidth]{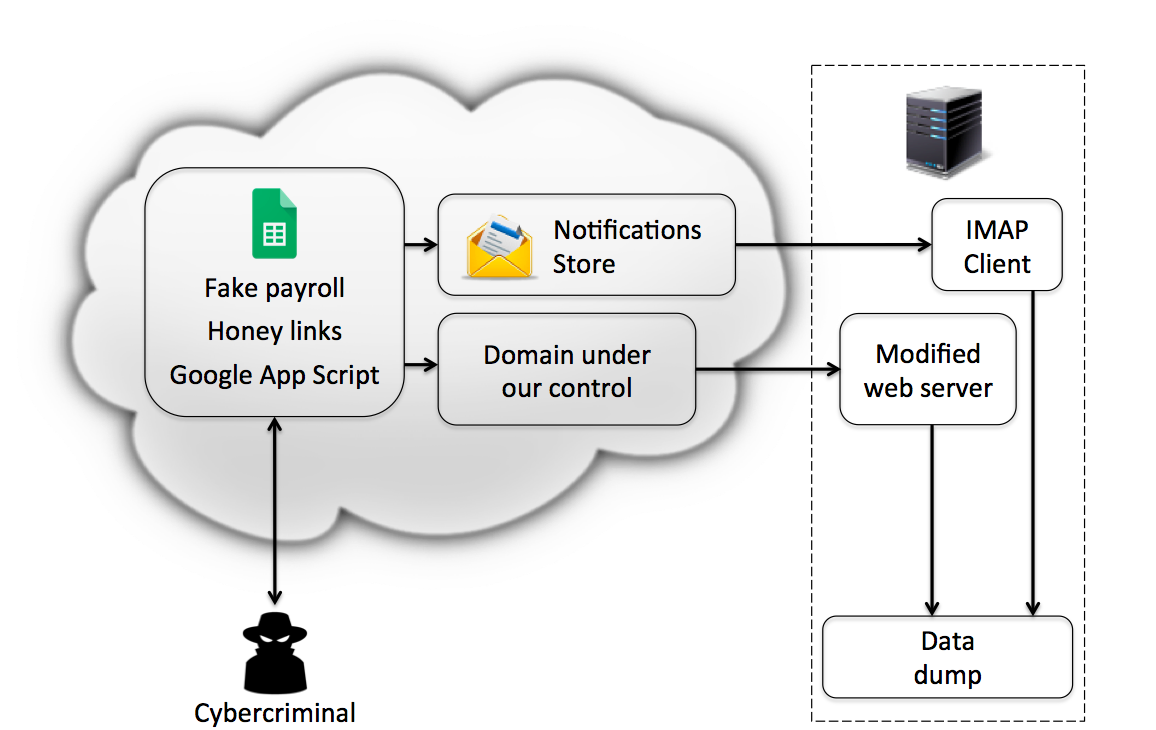}
  \caption{A simplified overview of our honeypot infrastructure. 
  Visitors interact with honey spreadsheets, and information 
  about their actions is sent to the webmail account that receives 
  notifications. If certain links in the spreadsheets are clicked, our modified 
  web server receives the generated HTTP requests, logs information 
  about the requests, and redirects them to 
  \texttt{google.com}~\cite{googledotcom}. We 
  also monitor clicks on all embedded links through Google Analytics, 
  since we used Google's URL shortener to obfuscate those links before 
  inserting them in the honey spreadsheets.}
  \label{fig:sys-overview}
\end{figure}

Our aim was to gain insight into what attackers do after gaining 
unauthorized access to leaked spreadsheets. 
To achieve this, we developed a system based on Google Apps
Script~\cite{googleappscript} to monitor events performed by users on Google
spreadsheets, such as accesses, modifications, and click on links. 
We then set up experiments in which we created five Google spreadsheets and
added honey URLs in them, some pointing to a domain under our 
control, others pointing to non-existent bank pages. Figure~\ref{fig:sys-overview} 
shows an overview of the honeypot infrastructure we set up, to monitor accesses 
to the honey spreadsheets. 
We leaked links pointing to the honey spreadsheets on 
\texttt{pastebin.com}~\cite{pastebin}, a popular 
paste site. The idea is to observe what captures the attention 
of attackers, and how they interact with the spreadsheets.

\subsection{Honey spreadsheets}
\label{subsec:honey-spreadsheets}

\begin{figure}[th]
  \centering
  \includegraphics[width=0.45\textwidth]{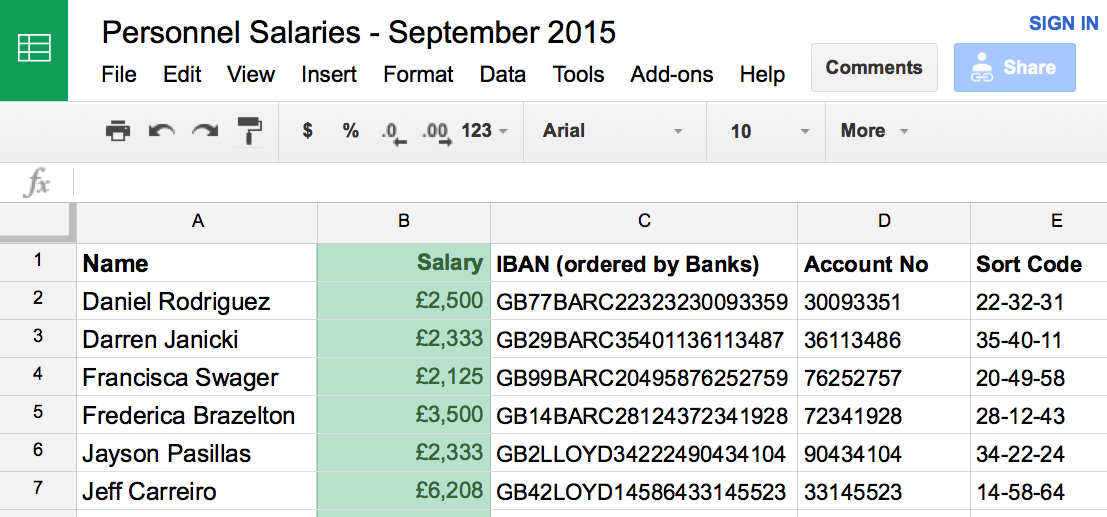}
  \caption{A partial view of a leaked honey spreadsheet, containing 
  fake payroll information. Note that shortened URLs are also in the same 
  spreadsheet, but not shown here.}
  \label{fig:partial-spreadsheet}
\end{figure}

We set up a Google account and created five spreadsheets in it. The 
spreadsheets were populated with randomly selected names and 
fake bank account details, including fake International Bank Account 
Numbers (IBANs), sort codes, and links masquerading as funds transfer 
pointers apparently connected to some of the fake 
accounts. Some of the links point to a domain under our control. The idea is 
to log attempts made by cybercriminals to transfer funds out of the 
non-existent honey accounts. One of the honey spreadsheets is shown in 
Figure~\ref{fig:partial-spreadsheet}.  

On creating a spreadsheet on the Google Docs platform, a unique link is 
generated by Google. Anyone with the link can thus access the spreadsheet, 
either with read-only permission or edit permission, depending on the 
access settings in the spreadsheet. We configured each spreadsheet to 
grant edit permission to anyone in possession of its unique link. In 
addition, the spreadsheets were instrumented with scripts to monitor activity, for instance, file open and edit events. The hidden 
scripts report those events back to us by sending email notifications to 
a webmail account under our control. The scripts were 
implemented in Google Apps Script~\cite{googleappscript}.

Our monitoring scripts periodically take snapshots of the honey 
spreadsheets, and send notification messages about those snapshots to a 
webmail account under our control. Information contained in the 
notifications include modified content, changes in spreadsheet 
structure, the date and time that the spreadsheet was last accessed, and 
date and time that the spreadsheet was last updated. We periodically 
retrieve and collate notifications from the webmail account, and parse 
them locally.

We set up an experiment involving spreadsheets whose URLs are posted on public
paste sites. This experiment shows the usefulness of our system, and
we believe that our approach will be helpful in future related 
work on the behavior of cybercriminals that compromise documents stored 
in the cloud. In Section~\ref{sec:discussion} we explore additional experiments
that could be performed to further understand the ecosystem around stolen online
documents.

\begin{figure}[th]
  \centering
  \includegraphics[width=0.25\textwidth]{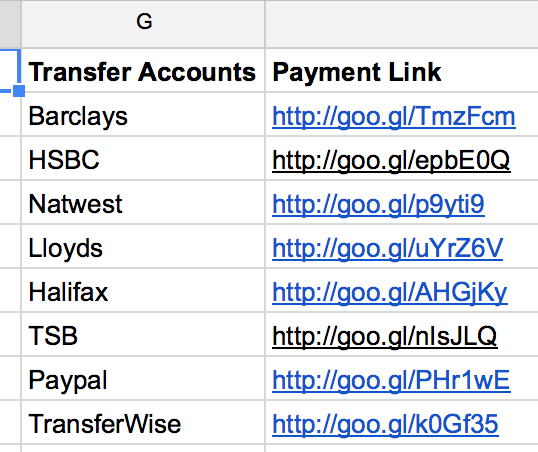}
  \caption{Honey links embedded in leaked honey spreadsheet. The URLs 
  were obfuscated using a URL shortener, and some of them point to a web server 
  under our control.}
  \label{fig:honey-links}
\end{figure}

\noindent \textbf{Honey links:} One of our goals was to 
observe if cybercriminals would attempt to 
observe or tamper directly with any of the honey bank accounts listed in 
the payroll. To achieve this, we did the following: In each honey 
spreadsheet, we inserted 9 URLs, 3 of them pointing to a domain under our 
control, and the remaining 6 pointing to non-existent bank pages. 
Each URL was shortened and obfuscated using 
\texttt{goo.gl}~\cite{googleurlshortener}, an online URL shortener. 
The idea is that all clicks on the embedded links would be logged by Google 
Analytics platform, which monitors URLs that 
were shortened using \texttt{goo.gl}~\cite{googleurlshortener}, 
and clicks on the 3 links pointing to our web server would be logged by the web server. 
Figure~\ref{fig:honey-links} shows the honey URLs we embedded in 
the honey spreadsheets. 

\subsection{Monitoring the spreadsheets}
\label{subsec:monitoring}
In addition to 
tracking changes in the honey spreadsheets using Google Apps Script, we 
incorporated honey links in each spreadsheet. As earlier stated, the 
idea is to track attempts by cybercriminals to stage further 
attacks on the honey accounts listed in the spreadsheets. 
Some of the embedded honey links point to a domain 
under our control, and clicks on those ones 
generate HTTP requests to a web server under our control. 
 
We configured the web server to redirect all incoming requests to 
\texttt{google.com}~\cite{googledotcom}, and log details such as the 
visitor's IP address, originating port number, HTTP header 
details, request paths, and timestamps of accesses. The purpose of the 
web server is to deflect the attention of visitors from our honeypot 
infrastructure, while logging accesses simultaneously.

We periodically run another script that connects to the webmail account 
that receives email notifications from the honey spreadsheets, to 
retrieve those notifications through the Internet Message Access Protocol (IMAP). 
This enables us to parse and analyse the notifications offline.

Links pointing to the honey spreadsheets were leaked on 
\texttt{pastebin.com}~\cite{pastebin}, to entice attackers to 
interact with the honey sheets. We set up two experiments: in the first one, we
pretended to be the legitimate owner of the spreadsheet, sharing it with a
colleague without realizing that this would make it accessible to the entire
Internet. In the second experiment we presented to be a criminal who got illicit
access to the spreadsheet and posted in on the paste sites to brag. Our experiments ran for 72 days. 

\subsection{Ethics}
\label{subsec:ethics}

Our experiments involve deceiving potential cybercriminals into interacting with 
fake spreadsheets containing fake financial information. We made sure that 
all information in the spreadsheets were not related to any real financial 
account. To mitigate the possibility of misuse of the online account 
hosting the honey spreadsheets, we did not leak credentials of the account, 
thus ensuring we remained in full control of the account throughout the 
experiments. Finally, to ensure that our experiments were run in an ethical manner, 
we sought and obtained permission from our institution's ethics committee, under project number 6521/001. 

\section{Results}

In this section, we discuss the activity observed in honey 
spreadsheets as part of our experiment. 
We also provide details of accesses to the modified web 
server under our control, and Google Analytics data we obtained 
from the shortened URLs we embedded in the spreadsheets.

\subsection{Overview}

Over the period of evaluation (72 days), we conducted two experiments with the 
honey sheets, with two main leak message themes. During the first 
experiment, we pretended to be a hacker that discovered some corporate 
financial information. For instance, we leaked messages like 
``leaked corporate payments'' and ``st0len payrolls'' along with 
unique links pointing to the spreadsheets. In the second experiment, we 
pretended to be a naive person sharing financial information 
with a colleague, and inadvertently leaking such information. In this 
case, the leak messages were of the form ``Bob, here is the spreadsheet 
with payrolls for September.'' The idea was to identify the leak message 
theme that would generate more interest in the honey sheets.

The first experiment, in which we pretended to be a hacker, was conducted 
from 23rd January 2016 till 8th March 2016 (46 days), during which 
links to the honey sheets were leaked twice daily. The second experiment, 
in which we pretended to naively share links with a colleague, was carried 
out from 9th March 2016 till 3rd April 2016 (26 days). Similarly, we 
shared unique links pointing to the honey sheets twice daily.

\subsection{Activity on honey spreadsheets}

For each leaked honey spreadsheet, we recorded file open events and 
file modification events. File open events are triggered when a honey 
spreadsheet is opened by visiting the unique URL pointing to it, 
while modification events are triggered when the structure of 
the spreadsheet is changed by the person accessing it. Modification 
events are also triggered when the contents of the spreadsheet 
are changed. We recorded snapshots of each spreadsheet every two 
hours, and compared those snapshots over time, to monitor the values 
that were modified in the spreadsheets.

In the first experiment, during which we pretended to be a hacker 
leaking financial information, we observed 112 file open events and 
17 modification events over a period of 46 days. 
In the second experiment, in which we posed as a naive user, 
we observed 53 file open events and 11 modification events 
over 26 days. We could not identify significant statistical 
differences between the two experiments with regards to 
file open and modification events. The dataset containing the details of the
open and modification events is available
at~\url{http://dx.doi.org/10.14324/000.ds.1502241}.

We acknowledge the possibility that some visitors may make copies 
of the spreadsheets for offline viewing and modification. This is one 
of the limiting factors in the experiments.

\noindent \textbf{Interesting modification events.} We observed some interesting
modification events among the 28 recorded during our experiments. 
In one of the honey spreadsheets, 
one of the bank account numbers was deleted, and the 
spreadsheet itself was rendered unusable. The attacker 
changed font sizes and colors, along with background colors in 
the spreadsheet, and replaced all the 
embedded honey links with the short URL~\url{https://goo.gl/ufniSo}. 
The short URL redirects to a Google search page for ``minions.'' 
In addition, the word `$\backslash$MINIONSXDDDD' was written on the spreadsheet, 
while an insult was entered in the spreadsheet. Interestingly, it 
appears the attacker did not click on any of the embedded honey links.
This case study shows that our honey spreadsheet infrastructure could be used to
study trolling, cyber bullying, and hate speech behavior on the web.

In another case, we observed a number of spreadsheet modifications that brought 
about no change in the content of the affected honey spreadsheets. 
A closer look revealed that the viewers expanded some columns, especially 
the ones containing fake payment links for funds transfer, presumably to have 
a better view of the URLs there.

\subsection{Activity on embedded honey links}

We observed 39 unique IP addresses that visited the 3 links 
pointing to our web server, and a total of 44 visits to 
those 3 links. We recorded accesses from 35 different 
countries, and a total of 174 clicks on all the 9 embedded 
honey links. Those clicks originated from the following browsers: 
Chrome, Firefox, Safari, and Samsung browsers, on devices running 
Windows, Linux, Macintosh, and Android operating systems.

\begin{figure*}[t]
  \centering
  \includegraphics[width=0.90\textwidth]{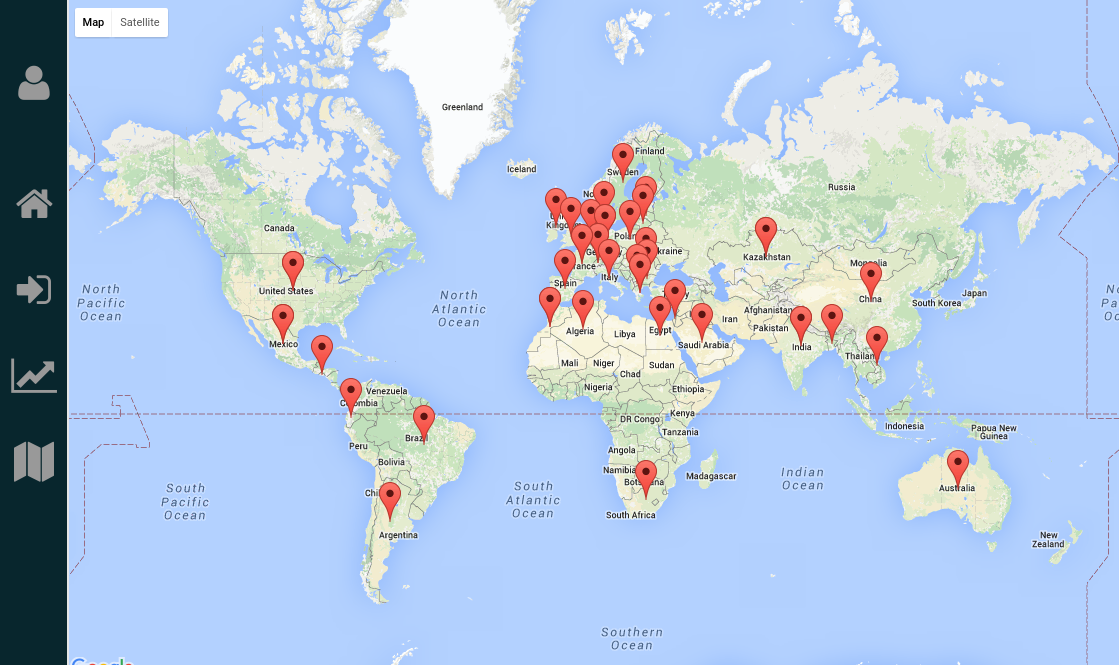}
  \caption{ Location of visits to the embedded honey links.} 
  \label{fig:world-map}
\end{figure*}

\noindent \textbf{Location of accesses.}
In order to have an idea of the locations of visitors 
accessing the honey spreadsheets, we extracted country information 
from Google Analytics dashboard, and also performed 
geolocation on IP addresses recorded by our web server. 
We plotted the locations on a world map, as shown in 
Figure~\ref{fig:world-map}.   

\section{Discussion}
\label{sec:discussion}

In this section, we mention the limitations that we encountered during the 
experiments, and discuss some additional experiments that could be performed
using our infrastructure to better understand the modus operandi of attackers
and cybercriminals illicitly accessing cloud documents.
  
\noindent \textbf{Limitations.} We were able to leak the spreadsheets only on 
a paste site (and not, for example, on an underground forum), 
implying that visitors accessing the spreadsheets 
are likely unsophisticated cybercriminals or just curious users. This is further 
emphasized by the fact that the visitors did not appear to make 
much effort to cover up their accesses. 

Attackers could make copies of the honey spreadsheets and explore 
them offline. We would be unable to monitor what they do with the 
spreadsheets offline, unless they click on the embedded honey links.

In Google spreadsheets, there are limitations on the functionality 
of Google Apps Script for users that are not logged in. Since visitors 
are not required to login before interacting with the spreadsheets, 
there are limits on the amount of information we can collect on the 
visitors, and also how much activity we can monitor in the 
spreadsheets. This is an intrinsic limitation of our monitoring 
infrastructure.

\noindent \textbf{Future work.} In the future, we plan to continue 
studying the modus operandi of cybercriminals interacting with 
compromised online documents, and the underground economy surrounding 
them. We intend to instrument more documents, and leak documents 
containing honey credentials linked to honey web services. The idea is 
that cybercriminals would attempt to log in to those honey web services 
using the honey credentials, and this hopefully would be a good way to 
distinguish between merely curious visitors and cybercriminals. As a result, 
leaking credentials in honey documents would give us broader visibility into 
their modus operandi.
We also plan to leak documents to more sites (for 
instance, through malware and underground forums), 
and study differences in accesses across different leak sites. 
We could modify the existing monitoring infrastructure we have 
developed to other online document storage platforms. 
We also plan to devise a taxonomy of attackers targeting 
online cloud documents. Finally, we intend to study online trolling 
and hate speech further, having discovered that our honey documents attract 
insults and destructive action (defacing).

\section{Related Work}
Honeypots have proved to be useful in cyber security studies, for instance, 
in the analysis of malware behaviour in controlled environments, tracking of
insider threats in organizations, and understanding cybercriminal schemes. 
In general, resources created with the sole purpose of being misused by
cybercriminals to enable researchers to learn their modus operandi are referred
to as \emph{honeytokens}. 
In this section, we compare our work with previous studies.

\noindent\textbf{Applications of honeytokens to security.} Liu et al.~\cite{liu2012many} studied the problem of privacy in 
Peer-to-Peer (P2P) networks, using honey files. Similarly, Salem 
et al.~\cite{salem2011decoy} and Spitzner~\cite{spitzner2003honeypots} 
proposed the use of decoy documents to track attackers, mitigate 
insider threats, and detect intruders~\cite{yuill2004honeyfiles}. 
Unlike those studies, we consider documents stored in the cloud instead 
of documents stored locally.
In 2012, Sobesto et al.~\cite{sobesto2012computer} studied the impact 
of computer configuration on computer-focused attacks, by deploying 
honeypot machines and allowing brute-force attackers to ``break'' into them 
via SSH. Unlike our work, they do not study compromised online documents.
Nikiforakis et al.~\cite{nikiforakis2011exposing} used honey files to expose
privacy leaks and abuse on file hosting services.
Kapravelos et al.~\cite{kapravelos2014hulk} used honey web pages to study
privacy invasive and malicious browser extensions on Google Chrome.
De Cristofaro et al.~\cite{de2014paying} created a Facebook page for a fake
company on Facebook, with the goal of studying the fake likes that it would
receive. Nikiforakis et al.~\cite{nikiforakis2014stranger} displayed fake
advertisements on ad-based URL shortening services, with the goal of exposing
their privacy and security issues.
A handful of papers~\cite{Stringhini:10:socialnet-spam,lee2010uncovering} deployed honey 
profiles on online social networks to identify accounts operated by spammers. 
Stringhini et al.~\cite{stringhini2014harvester} 
identified various actors in the email spam ecosystem, and their 
interrelationships, by leaking honey email addresses and monitoring 
spam emails received by them. 
In this paper we looked at honey Google Documents, with the goal of understanding the
activity of malicious users on such documents.

\noindent\textbf{Studies on compromised accounts and credentials.} Stone-Gross et al.~\cite{stone2011underground} 
investigated the \emph{Pushdo/Cutwail} botnet, and discovered that 
cybercriminals were selling online account information in underground 
forums. We used a similar approach to leak links pointing to honey 
spreadsheets on public paste sites.  
Bursztein et al.~\cite{bursztein2014handcrafted} studied 
account activity in Gmail accounts that were compromised mostly through 
phishing pages. In contrast to Bursztein et al.~\cite{bursztein2014handcrafted},
this paper focuses on activity in Google spreadsheets, rather than on Gmail
accounts.
Egele et al.~\cite{egelecompa} studied the problem of compromised online social
network accounts, developing behavioral models to detect and block these
compromises. We believe that the activity of
hijackers on honey documents will be significantly different than the one from
their legitimate owners. For these reasons, similar techniques to detect
compromises could be applied to the context of honey documents as well.

\section{Conclusion}

In this paper, we presented a honey spreadsheet system able to 
monitor the activity of cybercriminals interacting with 
compromised Google spreadsheets. We set up 5 honey 
spreadsheets, populated with fake financial data, 
and leaked them on a paste site. Prior to leaking them, we embed 
9 honey links in each spreadsheet, to track potential attempts 
to mount attacks against the fake financial accounts in the 
spreadsheets. We observed accesses to the spreadsheets, 
and attempts by visitors to access some honey links embedded in the 
spreadsheets, for a period of 72 days. We present 
preliminary results of the experiments and highlight how 
our method may be useful in studying malicious 
activity in online documents, in general.  

\section{Acknowledgments}
We wish to thank the anonymous reviewers for their comments. 
This work was funded by the H2020 RISE Marie Sklodowska Curie action (MSCA) grant number 691925, by
the EPSRC grant number N008448 and by a Google Faculty Research Award. Jeremiah
Onaolapo was supported by the Petroleum Technology Development
Fund (PTDF) of Nigeria.

\bibliographystyle{acm}
\bibliography{biblio}

\end{document}